\documentclass[12pt]{article}
\usepackage{amsmath,amssymb}
\usepackage{iftex}
\ifPDFTeX
  \usepackage[T1]{fontenc}
  \usepackage[utf8]{inputenc}
  \usepackage{textcomp} 
\else 
  \usepackage{unicode-math} 
  \defaultfontfeatures{Scale=MatchLowercase}
  \defaultfontfeatures[\rmfamily]{Ligatures=TeX,Scale=1}
\fi
\usepackage{lmodern}
\usepackage{multirow}
\usepackage{graphicx}

\usepackage{xcolor}
\usepackage[colorlinks=true,allcolors=blue]{hyperref}       
\usepackage[numbers]{natbib}
\newcommand{\R}{\mathbb{R}}
\newcommand{\vc}{\mathbf}

\usepackage[normalem]{ulem}

\usepackage[margin=1in]{geometry}
\usepackage{setspace}
\makeatletter
\def\maxwidth{\ifdim\Gin@nat@width>\linewidth\linewidth\else\Gin@nat@width\fi}
\def\maxheight{\ifdim\Gin@nat@height>\textheight\textheight\else\Gin@nat@height\fi}
\makeatother

\makeatletter
\def\fps@figure{htbp}
\makeatother
\ifLuaTeX
  \usepackage{selnolig}  
\fi
\IfFileExists{bookmark.sty}{\usepackage{bookmark}}{\usepackage{hyperref}}
\IfFileExists{xurl.sty}{\usepackage{xurl}}{} 
\hypersetup{
  hidelinks,
  pdfcreator={LaTeX via pandoc}}

\title{\textbf{Physics-Informed Latent Space Dynamics Identification for Time-Dependent NLTE Atomic Kinetics}}
\author{%
\parbox{0.95\textwidth}{\centering
Jeongwoo Nam\textsuperscript{1,2},
William Anderson\textsuperscript{3},
Youngsoo Choi\textsuperscript{3},
Hai P. Le\textsuperscript{3},
Mark E. Foord\textsuperscript{3},
Byoung Ick Cho\textsuperscript{1,2,$\dagger$},
Haewon Jeong\textsuperscript{4,$\ddagger$},
Min Sang Cho\textsuperscript{3,*}
\\[0.5ex]
\small
\emph{\textsuperscript{1} Department of Physics and Photon Science, Gwangju Institute of Science and Technology (GIST), Gwangju 61005, Republic of Korea}\\
\emph{\textsuperscript{2} Center for Relativistic Laser Science, Institute for Basic Science (IBS), GIST, Gwangju 61005, Republic of Korea}\\
\emph{\textsuperscript{3} Lawrence Livermore National Laboratory, 7000 East Avenue, Livermore, California 94550, United States of America}\\
\emph{\textsuperscript{4} Department of Electrical and Computer Engineering, University of California, Santa Barbara (UCSB), California 93106, United States of America}\\[0.5ex]
\small
\textsuperscript{*}\texttt{cho28@llnl.gov},
\textsuperscript{$\dagger$}\texttt{bicho@gist.ac.kr},
\textsuperscript{$\ddagger$}\texttt{haewon@ece.ucsb.edu}
}}
\date{}

\begin{document}

\maketitle

\begin{abstract}
Non-local thermodynamic equilibrium (NLTE) calculations remain a major computational bottleneck in radiation--hydrodynamics, while most existing machine-learning surrogates treat NLTE as a static input--output mapping rather than a kinetic evolution problem. Here, we present a physics-informed Latent Space Dynamics Identification (pLaSDI) framework specifically designed for NLTE atomic kinetics, which captures the time-dependent atomic kinetics of non-equilibrium plasmas through an explicit reduced governing equation. To ensure the physical reliability of the reduced model, we impose physics-informed loss terms that enforce macroscopic consistency, dynamical stability, and convergence to the correct steady state during long-time integration. Applied to tin NLTE population data generated along hydrodynamically modeled temperature--density trajectories relevant to extreme ultraviolet (EUV) lithography plasmas, the model accurately reproduces charge-state evolution and mean charge state with errors below 2\%, achieves speedups of approximately $5\times10^{4}$--$10^{5}$, and remains stable outside the training trajectories by converging toward physically admissible states and the correct steady-state solution under fixed plasma conditions. These results show that careful physics-informed design of the latent dynamics, rather than data fitting alone, is essential for constructing fast, stable, and physically reliable extrapolative surrogates for time-dependent NLTE kinetics.
\end{abstract}

\newpage

\section{Introduction}

Non-local thermodynamic equilibrium (NLTE) atomic kinetics plays a central role in radiation–hydrodynamics simulations by providing self-consistent closures for opacity, emissivity, and equation-of-state quantities \cite{scott2001cretin, FRANK2022100998, scott2022using}. In radiation–hydrodynamics simulations of a wide range of laser–plasma experiments, such as extreme ultraviolet (EUV) source development, NLTE calculations must be repeatedly performed at every hydrodynamic time step \cite{scott2022rad_hydro_lpp_euv}. This creates a severe computational bottleneck, often dominating the total simulation cost and limiting the practicality of large-scale ensemble simulations required for design optimization and uncertainty quantification. Consequently, the development of fast and reliable NLTE models is not merely desirable, but essential for enabling next-generation predictive workflows.

Recent studies on machine-learning (ML)-based NLTE modeling have shown that surrogate methods can substantially reduce the computational cost of NLTE calculations. In particular, because NLTE models often serve as sub-physics modules that provide radiative properties in radiation--hydrodynamics simulations, many existing approaches have formulated NLTE as a direct input--output mapping problem from plasma conditions to spectra or other radiative observables. This perspective has made it relatively straightforward to deploy neural-network-based surrogates, and early studies established the basic feasibility of this strategy \cite{kluth2020deep}. Subsequent developments, including transfer learning \cite{vanderwal2022neural, vander2023transfer}, multi-network architectures \cite{schaeuble2025deep}, and physics-informed variable transformations \cite{cho2025PRR}, further improved the ability to reproduce high-fidelity opacity spectra and related NLTE observables even from limited expensive data. More recently, motivated by the growing importance of time-dependent NLTE calculations \cite{bishel2023ionization, cho2025reduced, cho2025ionization}, several studies have begun to move beyond static regression by introducing physics-assisted autoencoders and learned reduced dynamics to represent stiff collisional-radiative evolution in latent space \cite{xie2024latent, zhang2025deep}. Despite this progress, however, the long-time physical reliability of ML-NLTE surrogates remains insufficiently addressed, particularly with respect to stability, asymptotic consistency, and behavior outside the training regime. In particular, many current approaches still retain the character of black-box or weakly constrained surrogates, which limits their extrapolative robustness as NLTE physics continues to evolve, for example through changes in ionization potential depression models \cite{ciricosta2012direct} or the inclusion of non-Maxwellian electron distributions \cite{le2019influence, cho2024PRE}.

To address these limitations, we reformulate NLTE surrogate modeling not merely as a time-dependent prediction problem, but as a physics-constrained kinetics-learning problem. Specifically, we develop a physics-informed Latent Space Dynamics Identification (pLaSDI) framework tailored to NLTE atomic kinetics, in which the evolution of non-equilibrium plasmas is represented through an explicit reduced governing equation in a learned latent space. Our approach builds on the LaSDI framework, which combines dimensionality reduction and symbolic regression to construct reduced-order models of dynamical systems \cite{fries2022lasdi, he2023glasdi, bonneville2024gplasdi, bonneville2024comprehensive}. By identifying an explicit latent dynamical system, rather than a black-box input--output map, the framework enables physical constraints to be imposed directly on the learned operator and allows its long-time behavior to be analyzed \emph{a priori}. We also incorporate physics-informed loss functions that enforce population conservation, charge-state consistency, dynamical stability, and correct steady-state behavior, while preserving the first-order structure of the original rate equations in the latent space. As a result, the model achieves substantial computational acceleration while maintaining high fidelity in key observables, including the charge-state distribution and mean charge state, together with improved robustness, extrapolative capability, and convergence toward physically admissible states outside the training regime. This work establishes a pathway toward scalable, physically reliable NLTE modeling for EUV lithography and laser--plasma science, including applications in fusion energy.

Methodologically, the present work builds on and extends a broader class of learned latent dynamical models that seek low-dimensional representations with simple, approximately linear evolution. 
In particular, VAMPnets \cite{Mardt2017} and Koopman-inspired deep-learning approaches \cite{lusch2018, takeishi2017} have shown that nonlinear kinetics can often be represented effectively in learned latent spaces governed by linear or near-linear dynamics. 
While these ideas are conceptually aligned with our approach, TD-NLTE calculations introduce additional requirements: the dynamics are explicitly driven by time-dependent plasma conditions, the latent variables must decode back to physically meaningful atomic population states, and the resulting predictions must recover correct charge-state distributions and average ionization. 
In addition, the learned dynamics must remain stable and converge to the correct steady state during long-time integration. 
These requirements motivate the physics-informed constraints introduced in this work.

The remainder of this paper is structured as follows: Section~\ref{sec:nlte_modeling} reviews NLTE atomic kinetics and introduces the population state variables, macroscopic observables, and key features of the governing kinetic equations and dataset. Section \ref{sec:lasdi} introduces baseline LaSDI for TD‑NLTE kinetics, including encoding/decoding and latent dynamics identification. Section \ref{sec:plasdi} presents our physics‑informed LaSDI method, detailing the macroscopic consistency constraints, the differentiable Hurwitz stability loss, and the steady‑state consistency loss, and summarizes the full training objective. Section \ref{sec:problem_setup_data} describes the problem setup and data generation procedure. Section \ref{sec:result} reports numerical experiments, metrics, and ablation studies that isolate the impact of each physics‑informed term, along with accuracy, stability, and efficiency results. Section \ref{sec:conclusion} concludes the paper with a discussion of the main findings, implications, limitations, and possible extensions to broader plasma conditions and multi-species problems.

\section{NLTE Kinetic Modeling}
\label{sec:nlte_modeling}

\subsection{Atomic State Population and Rate Equations}

NLTE atomic kinetics evolve a high-dimensional population vector that resolves internal structure within each charge state. Let $n_i(t)$ denote the population of atomic state $i$, where the index $i$ may represent a fine-structure level, a configuration, or a superlevel depending on the chosen atomic model. Collecting all states yields a vector

\begin{equation}
\mathbf{n}(t) = [n_1(t), \ldots, n_N(t)]^{T},
\end{equation}
with $N$ ranging from $\mathcal{O}(10^2)$ to $\mathcal{O}(10^6)$ for medium--high-$Z$ elements when detailed configuration resolution is retained \cite{scott2022using,Andrievsky2017Mnras}. A standard NLTE formulation expresses the time evolution as a linear, population-conserving rate system,

\begin{equation}
\frac{d n_i}{dt} = \sum_{k \ne i} R_{ik}(T,\rho,\mathbf{j})\, n_k-R_{ki}(T,\rho,\mathbf{j})\, n_i,
\label{eq:rate_FOM}
\end{equation}
where $T(t)$ and $\rho(t)$ are the scalar electron temperature and density, and $\mathbf{j}(t)$ denotes the radiation field, treated here as a vector quantity, e.g., a discretized spectral intensity $j_\nu$ \cite{ralchenko2016modern, cho2020opacity}. The rate matrix ${R}$ is assembled from atomic collisional and radiative processes and is typically sparse and structured. Off-diagonal terms represent transitions $k \rightarrow i$, while diagonal terms enforce the total outflow from state $i$. The transition set includes bound--bound excitation and de-excitation (electron-impact and radiative), bound--free ionization and recombination (including radiative, dielectronic, and three-body channels), and, when relevant, photo-processes driven by $\mathbf{j}(t)$ (e.g., photoexcitation and photoionization), as well as additional physics such as autoionization. Two aspects make the direct numerical solution of~\eqref{eq:rate_FOM} computationally expensive in NLTE physics workflows. First, the wide range of transition rates produces a stiff ODE system, so stable time advancement generally requires implicit integration with repeated solution of large coupled linear systems. Second, assembling or updating the rate matrix $R$ is itself expensive because the underlying collisional and radiative coefficients depend on the evolving plasma conditions and must be repeatedly reevaluated as $T(t)$ and $\rho(t)$ change, which can itself involve substantial atomic-rate calculations. Together, these costs motivate the use of reduced-order models.


For controlled testing and method development, we follow the stiff rate-equation benchmark setting of the referenced latent-dynamics study. Note that we adopt a simplified input dependence in which the driving parameters are limited to $T(t)$ and $\rho(t)$. In particular, the sample problems considered below assume an optically thin plasma, so radiative-transport feedback through $\mathbf{j}(t)$ is neglected or treated as fixed, and the effective operator is written as ${R}(T(t),\rho(t))$. This choice isolates the stiff multi-timescale kinetics and long-time relaxation behavior that are most relevant to surrogate stability and steady-state consistency. Extending the surrogate to radiation-coupled problems with explicit $\mathbf{j}(t)$ dependence is a natural next step.

\subsection{Macroscopic Observables and Conservation Constraints}

In addition to the microscopic populations $\mathbf{n}(t)$, several macroscopic quantities derived from them play an important role in NLTE modeling. Many multiphysics couplings and experimental interpretations depend more directly on these aggregate observables than on individual level populations. In this work, we focus on three such quantities that are both physically meaningful and directly testable.

\begin{enumerate}
    \item[(i)] \textbf{Population conservation.} For a single atomic species with total number density $n_A$ (or when using normalized fractional populations), the populations must satisfy a conservation law. In a fractional representation,
    \begin{equation}
    \sum_{Z,j} n_{Z,j}(t) = 1
    \qquad \text{for all } t,
    \label{eq:conservation}
    \end{equation}
    whereas in a density representation,
    \begin{equation}
    \sum_{Z,j} n_{Z,j}(t) = n_A
    \qquad \text{for all } t.
    \end{equation}
    Preserving this constraint is necessary for physical consistency and stable coupling to hydrodynamics.

    \item[(ii)] \textbf{Charge-state distribution (CSD).} The charge-state population, summed over all internal levels belonging to charge state $Z$, is
    \begin{equation}
    p(Z,t) = \sum_{j \in \mathcal{J}_Z} n_{Z,j}(t),
    \label{eq:csd}
    \end{equation}
    where $\mathcal{J}_Z$ denotes the set of internal levels associated with charge state $Z$.

    \item[(iii)] \textbf{Mean charge state.} The mean charge state is
    \begin{equation}
    \bar{q}(t) = \sum_{q=0}^{q_{\max}} q\, p(q,t).
    \label{eq:zbar}
    \end{equation}
\end{enumerate}

These quantities provide compact and physically interpretable summaries of the underlying population dynamics. They are therefore useful not only for analyzing NLTE behavior, but also for assessing whether a reduced model preserves the physically important structure of the full solution.

\subsection{Steady State and Asymptotic Convergence}

Under fixed plasma conditions, the steady state is defined by the vanishing of the time derivative of the population vector, $\frac{d\mathbf{n}}{dt} = 0$. Substituting this condition into the rate equation gives
\begin{equation}
R(T,\rho)\,\mathbf{n}^{\ast} = 0,
\end{equation}
subject to normalization constraints such as total population conservation. In general, this steady state is not a thermodynamic equilibrium, but an NLTE equilibrium established by the balance of collisional and radiative processes.
Two intrinsic dynamical properties of the NLTE rate equations are particularly important for surrogate modeling:
\begin{enumerate}
    \item[(i)] \textbf{Steady-state existence.} For fixed plasma conditions, the system admits a stationary solution $\mathbf{n}^{\ast}$ determined by the balance of the underlying atomic processes. A physically meaningful surrogate should preserve the existence of this steady state under the same conditions.

    \item[(ii)] \textbf{Asymptotic convergence.} NLTE systems exhibit relaxation toward the steady state as time progresses. Owing to the dissipative structure of the rate operator, physically admissible solutions converge toward $\mathbf{n}^{\ast}$ as $t \to \infty$, and the convergence behavior is governed by the eigenvalue spectrum of $R$. 
\end{enumerate}

The associated relaxation time scales can span multiple orders of magnitude. It makes NLTE systems inherently stiff and renders their long-time behavior highly sensitive to small errors in the learned dynamical operator. As a result, a surrogate trained only to minimize short-time data mismatch may reproduce transient trajectories within the training window, yet still fail to converge to the correct steady state or exhibit unstable growth during extrapolation. In time-dependent plasmas, where T and $\rho$ evolve continuously, the system may be viewed as relaxing toward a sequence of instantaneous steady states. Therefore, a physically reliable surrogate must capture not only short-time dynamics but also the correct asymptotic behavior under fixed conditions. This motivates the introduction of steady-state and convergence-aware constraints in the learning objective. By enforcing that the learned dynamics admit the correct steady state and converge stably toward it, the surrogate can be made to preserve the essential properties of NLTE kinetics even beyond the training time horizon.

\subsection{Atomic Population Scaling and w-Scaled Representation}
\label{sec:wscaled}
The atomic population values $\mathbf{n}$ typically span an extremely wide dynamic range, approximately from $10^{-50}$ to $1$. If the original-scale population values are directly used as inputs to the autoencoder and the reconstruction loss is defined only in the original population space, population values below the converged loss scale contribute negligibly to the gradient and are effectively invisible to the optimizer. As a result, reconstruction accuracy in the small-population regime can degrade substantially. Although these small populations are negligible in absolute magnitude, they can still affect derived radiative quantities such as spectra, opacity, and emissivity, and therefore should not be ignored. To alleviate this imbalance, we adopt a w-scaled population representation, defined through a logarithmic transformation followed by min--max normalization:
\begin{equation}
\mathbf{w}' = 1 - \log(\mathbf{n}),
\end{equation}
\begin{equation}
\mathbf{w} = \mathrm{MinMax}(\mathbf{w}'), \qquad \mathbf{w} \in [0,1].
\end{equation}
Under this transformation, extremely small population values in the original scale are mapped to values close to 1 in the w-space, while large population values are compressed toward values close to 0. More generally, preprocessing transformations are commonly employed in ML-based NLTE surrogate modeling when the target variables span a wide dynamic range. Their primary role is to improve numerical balance during training and to enhance the representation of small-magnitude states in the loss function \cite{cho2025PRR,xie2024latent,VANDERWAL2022ML}. In this sense, the w-scaled representation serves as a practical preprocessing step that improves the numerical conditioning of the reconstruction problem.

\section{LaSDI Framework}
\label{sec:lasdi}
Conventional linear reduced-order methods often leverage the proper orthogonal decomposition (POD) to project high-dimensional states onto a fixed linear subspace. 
However, NLTE population dynamics are strongly nonlinear: atomic populations shift abruptly across charge states as plasma conditions change, and the resulting manifold structure cannot be adequately captured by a linear basis. This leads to significant errors during long-time integration.
Additionally, forming the reaction matrix $R$ at every timestep is computationally expensive, limiting potential speedup from projecting directly onto the governing ODE~\eqref{eq:rate_FOM}.

To overcome these limitations, we employ Latent Space Dynamics Identification (LaSDI) \cite{fries2022lasdi, bonneville2024gplasdi}. 
LaSDI provides a general framework for constructing reduced-order models of parameterized dynamical systems by combining a low-dimensional representation of the state with an identified model for the associated reduced dynamics. 
For the NLTE problems considered in this work, we adopt a specific realization of this framework using an autoencoder-based nonlinear compression and an explicit ODE model for the latent-space dynamics. 
We further incorporate physics-informed constraints, including stability-related constraints, in the reduced model for the NLTE problems considered in this work.
For a new input parameter, the initial condition is mapped to the reduced representation, evolved according to the learned reduced dynamics, and then reconstructed in the full state space. 
Detailed descriptions of LaSDI and its variants can be found in Refs.~\cite{fries2022lasdi,he2023glasdi, bonneville2024gplasdi, bonneville2024comprehensive}.

Applying LaSDI to NLTE kinetics presents two challenges that, to our knowledge, have not been addressed in prior LaSDI variants.
First, the temperature and density inputs are time-dependent rather than fixed scalars, so standard interpolation of trajectory-specific ODE coefficients is not directly applicable.
To handle time-dependent input parameters, we replace trajectory-wise coefficient interpolation by using Dynamics Mode Decomposition with control (DMDc) \cite{proctor2016}, which allows temperature and density to enter the latent dynamics directly as control variables.
The second challenge is that our model must capture the long-term steady-state dynamics, beyond just the training time window.
To promote stable long-time dynamics, we impose a Hurwitz stability condition during training.
We describe our approach in detail below.

\subsection{Latent Space Dynamics Identification}

LaSDI is a purely data-driven, reduced-order modeling framework.
Here, we compress the atomic states with a nonlinear autoencoder. 
Our autoencoder consists of an encoder $\mathcal{G}_{\mathrm{enc}}$ which maps the high-dimensional population vector $\mathbf{n} \in \mathbb{R}^{N}$ to a low-dimensional latent state $\mathbf{z} \in \mathbb{R}^{N_z}$ ($N_z \ll N$), and the decoder $\mathcal{G}_{\mathrm{dec}}$ reconstructs $\mathbf{n}$ from $\mathbf{z}$. 

More precisely, for the $k$-th pair of temperature and density $( T^{(k)}(t), \ \rho^{(k)}(t) )$, we consider snapshots of the atomic states $\vc n^{(k)}$ taken at uniform times $t_i, \ i = 0, 1, ..., N_t$.
We then form these snapshots into a training data matrix $X^{(k)} = [\vc n^{(k)}(0), \vc n^{(k)}(t_1), ... , \vc n^{(k)}(t_{N_t}) ]^\top \in \R^{(N_t + 1) \times N}$. 
Concatenating the snapshots from each of our $N_{\mathrm{Train}}$ training trajectories, we obtain the training data tensor
\begin{equation}
    \mathbf{X} := [  X^{(1)},  X^{(2)}, ...,  X^{(N_{\mathrm{Train}})} ]\in \R^{N_{\mathrm{Train}} \times (N_t + 1) \times N}.
\end{equation}
The encoder and decoder then attempt to minimize the training reconstruction loss \begin{equation}
    \mathcal L_{\text{AE}} ( \pmb \theta_\text{enc} , \pmb \theta_\text{dec} ) = \| \mathbf{X} - \mathcal{G}_{\mathrm{dec}}( \mathcal{G}_{\mathrm{enc}}( \vc X ))\|^2.
\end{equation}  
Here  $\| \cdot \|$ is the Frobenius norm and $\pmb \theta_\text{enc}, \ \pmb \theta_\text{dec}$ are model parameters of the encoder and decoder.

After forming our training data tensor, we can also define our compressed training data tensor, consisting of snapshots in the latent space
\begin{equation}
    \mathbf{Z} := \mathcal{G}_{\text{enc}} ( \mathbf{X} ) \in \R^{N_{\mathrm{Train}} \times (N_t + 1) \times N_z}.
    \label{eq:compressed_tensor}
\end{equation}

\begin{figure}[htbp]
    \centering
    \includegraphics[width=\linewidth]{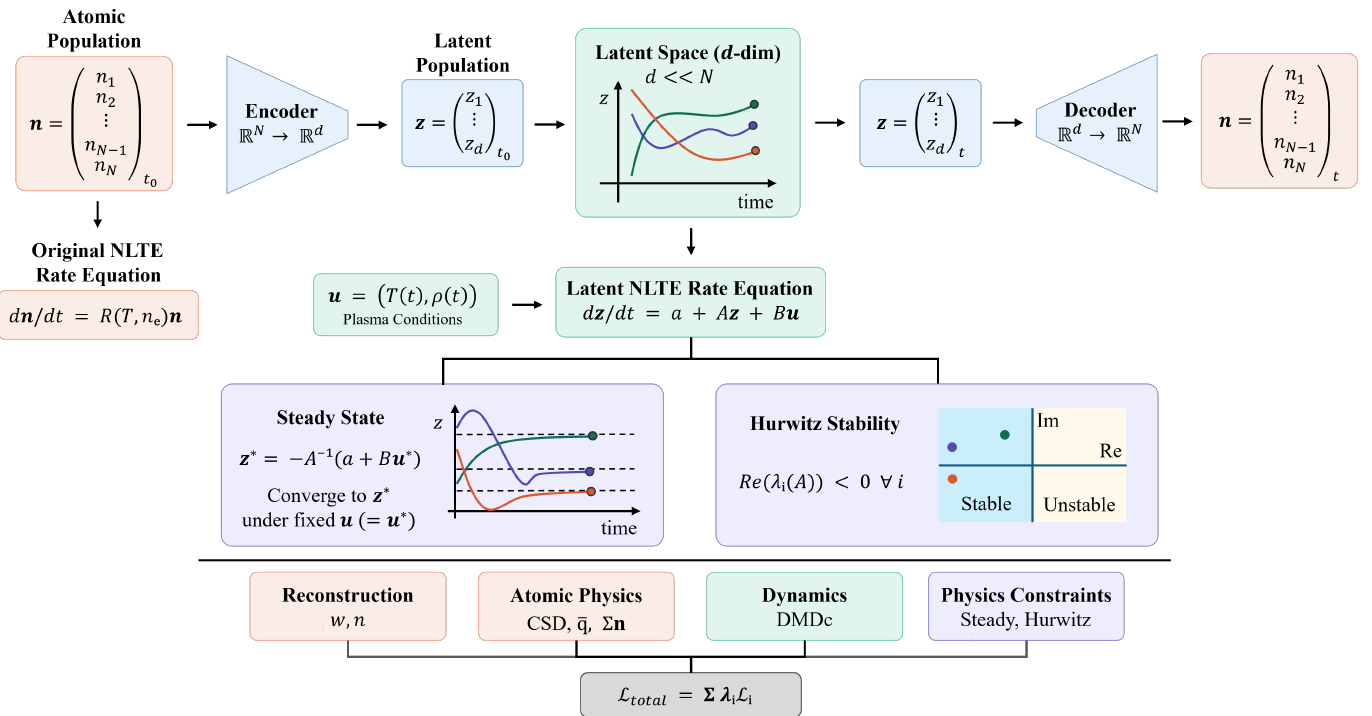}
    \caption{Overview of the physics-informed LaSDI (pLaSDI) framework for NLTE atomic kinetics. An autoencoder compresses the high-dimensional population vector $\mathbf{n} \in \mathbb{R}^N$ into a latent state $\mathbf{z} \in \mathbb{R}^d$ ($d \ll N$), where dynamics are governed by an explicit ODE identified via DMDc with plasma conditions $\mathbf{u}(t) = (T(t), \rho(t))$ as control variables. Physics-informed constraints enforce Hurwitz stability and steady-state consistency of the latent dynamics. The total loss combines reconstruction, macroscopic physics, dynamics identification, and physics constraint terms.}
    \label{fig:plasdi_overview}
\end{figure}

By absorbing the strong nonlinearity of NLTE kinetics into the encoder--decoder pair, the temporal evolution in the latent space can be approximated by a relatively simple dynamical system.
Most variants of LaSDI identify the dynamical system of the latent space via symbolic regression using Sparse Identification of Nonlinear Dynamics (SINDy) \cite{brunton2016sindy}, although any dynamics identification algorithm can be used such as Dynamic Mode Decomposition \cite{proctor2016, schmid2010}, GFINNs \cite{Zhang2022}, or Neural ODE \cite{chen2018, longhi2026}.
Figure~\ref{fig:plasdi_overview} provides an overview of the complete framework, including the autoencoder, latent dynamics identification, and physics-informed constraints.

\subsection{DMDc: Dynamic Mode Decomposition with Control}

Previous studies using the LaSDI framework usually approximate temporal evolution of each latent state $\mathbf{z}$ with a first-order linear ordinary differential equation

\begin{equation}
\frac{d\mathbf{z}}{dt} = \mathbf{a} + A\mathbf{z},
\label{eq:sindy}
\end{equation}
where $\mathbf{a}\in \R^{N_z}$ is a coefficient vector and $A \in \R^{N_z \times N_z}$ is a coefficient matrix identified by SINDy. In the conventional LaSDI approach, distinct coefficients $\mathbf{a}^{(k)}$ and ${A}^{(k)}$ would be identified for each training trajectory corresponding to a fixed temperature--density condition $( T^{(k)}, \ \rho^{(k)} )$. For unseen input parameters, they then interpolate new ODE coefficients using the dynamical systems fit to the training data, for example using Gaussian Processes  \cite{bonneville2024gplasdi} or a $k$-nearest neighbors approach \cite{he2023glasdi}.

However, it is unclear how to implement these interpolation schemes when using time-dependent input parameters, such as our temperature and density, rather than fixed scalar parameters.
Since the local coefficients are indexed by fixed $(T(t), \rho(t))$ conditions, there is no clear prescription for selecting or interpolating coefficients along a continuously evolving trajectory---particularly when the trajectory passes through regions of the parameter space not covered by the training set. 

To address this, we adopt an approach inspired by Dynamic Mode Decomposition with Control (DMDc) \cite{proctor2016} and other control methods \cite{fasel2021sindyc}, in which temperature and density are explicitly incorporated as control variables.
Our latent dynamics are then approximated by the dynamical system
\begin{equation}
\frac{d\mathbf{z}}{dt} = \mathbf{a} + {A}\mathbf{z} + {B}\mathbf{u},
\label{eq:DMDc}
\end{equation}
where $\mathbf{u}(t) = [\widetilde{T}(t),\, \widetilde{\rho}(t)]^\top$ denotes the control vector composed of the w-scaled temperature and density, obtained by applying the scaling transformation described in Section~\ref{sec:wscaled} to each variable independently, and ${B} \in \R^{N_z \times 2}$ is the associated coefficient matrix. 
We adopt a global coefficient formulation and learn a single set of coefficients $\{\mathbf{a}, {A}, {B}\}$ across all trajectories, providing a consistent dynamical model applicable to continuously varying conditions.
Unlike classical SINDy with control \cite{fasel2021sindyc}, we do not construct a library of candidate nonlinear functions or impose sparsity on the coefficients. 
Our formulation also differs from canonical DMDc in that it is posed directly in compressed latent coordinates and in continuous time, rather than as a discrete-time snapshot model in the original state variables. 
Although our model differs slightly from DMDc, the core algorithm is similar we shall refer to our approach to dynamics identification as DMDc regardless.

To define a loss function involving DMDc, we first concatenate snapshots of the control vector into a matrix $U^{(k)} = [\vc u^{(k)}(0), \vc u^{(k)}(t_1), ... , \vc u^{(k)}(t_{N_t}) ]^\top \in \R^{(N_t + 1) \times 2}$. 
We then form a tensor by concatenating the controls corresponding to each training trajectory
\begin{equation}
    \mathbf{U} = [  U^{(1)},  U^{(2)}, ...,  U^{(N_{\mathrm{Train}})} ]\in \R^{N_{\mathrm{Train}} \times (N_t + 1) \times 2}.
\end{equation}
To determine our DMDc coefficients, we minimize the loss
\begin{equation}
\mathcal{L}_{\mathrm{DMDc}} (\vc a, A, B) = \lambda_{\mathrm{DMDc}} \sum_{i = 1}^{N_\mathrm{Train}} \sum_{j = 0}^{N_t}\left\| \dot{\vc z}^{(i)}(t_j) - (\mathbf{a} + {A}\mathbf{z}^{(i)}(t_j) + {B}\mathbf{u}^{(i)}(t_j)) \right\|^{2},
\label{eq:loss_sindy}
\end{equation}
where $\lambda_{\mathrm{DMDc}}$ is a weight hyperparameter. 
In this work, we approximate $\dot{\vc z}$ using a centered difference, although a weak formulation of our approach can also be applied with LaSDI \cite{messenger2021weak}.
Our DMDc loss constrains the time derivative through an explicit differential equation, thereby enforcing temporal continuity in the latent space, which is not guaranteed by only minimizing the autoencoder loss.

In practice, the trained model requires only the coefficients $\{\mathbf{a}, {A}, {B}\}$, the autoencoder, an initial population, and the time-dependent control trajectories $\mathbf{u}(t)$---directly analogous to the temperature--density history files used by conventional SCFLY code \cite{chung2005flychk, Chung2008FLYCHKManual}. Most importantly, because the latent dynamics take the form of an explicit ODE, the stability and steady-state properties of the system can be analyzed mathematically---a capability that motivates the physics-informed constraints introduced in the following section.

\section{pLaSDI: Physics-informed LaSDI}
\label{sec:plasdi}
%
\subsection{NLTE Kinetic Constraints}
\label{sec:nlte_kinetic}

The latent dynamics learned through DMDc enforce continuous temporal evolution and suppress nonphysical oscillations; however, temporal continuity alone does not guarantee stability or convergence to a physically meaningful steady state. 
In time-dependent NLTE problems, even under fixed temperature and density, dynamics learned by LaSDI models may diverge or converge to incorrect equilibria during long-time integration. 
This limitation arises because the dynamics identification in LaSDI is based on a least-squares regression of time derivatives, which minimizes local fitting errors over the training window but does not impose constraints on the physical validity or asymptotic behavior of the resulting dynamical system.

A key advantage of the LaSDI framework is that it does not treat NLTE kinetics as a black-box mapping, but instead explicitly identifies a governing dynamical system in the latent space. By constructing the latent dynamics in the form of a first-order differential equation—designed to resemble the structure of the original NLTE rate equations—the model retains a direct connection to the underlying physics. This structural alignment enables us to impose NLTE-specific physical constraints, such as stability and steady-state convergence, directly on the learned latent operator. In contrast to conventional black-box models, where such constraints are difficult to enforce, the explicit and interpretable form of the latent dynamics allows us to analyze and control the long-time behavior \emph{a priori}. This provides a principled foundation for introducing physics-informed constraints, which are essential for ensuring reliability in long-time NLTE simulations.

\subsubsection{Hurwitz Stability Condition}

In our model, the latent dynamics are designed to be linear~\eqref{eq:DMDc}. Under fixed plasma conditions, \emph{i.e.}, $\mathbf{u}=\mathbf{u}^{\ast}$, the governing equation reduces to
\begin{equation}
\frac{d\mathbf{z}}{dt} = {A}\mathbf{z} + \mathbf{b},
\end{equation}
where $\mathbf{b} = \mathbf{a} + B\mathbf{u}^{\ast}$ is constant. The corresponding solution can be written as
\begin{equation}
\mathbf{z}(t) = {C}e^{{A}t} - {A}^{-1}\mathbf{b},
\end{equation}
where ${C}\in \R^{N_z \times N_z}$ is determined by the initial condition. 
Thus, the long-time behavior of the latent dynamics are governed by the matrix exponential $e^{{A}t}$. For the latent trajectory to remain bounded and converge to a finite equilibrium, all eigenvalues of ${A}$ must have strictly negative real parts. This is precisely the Hurwitz stability condition for a linear dynamical system. 
If this condition is violated, the latent trajectory may diverge during time integration, resulting in nonphysical population predictions. Therefore, within the present reduced-order formulation, stability can be assessed \emph{a priori} from the spectrum of ${A}$.

Motivated by this observation, we introduce a stability penalty during training to enforce the Hurwitz condition.
A natural choice is to penalize the eigenvalues of $A$ directly through minimizing the loss
\begin{equation}
\mathcal L_{\mathrm{stability}}(A) =
\mathrm{ReLU}\!\left(
\max \bigl( \mathrm{Re}(\lambda({A})) \bigr)
\right),
\end{equation}
which penalizes any eigenvalue of ${A}$ with a positive real part. However, because ${A}$ may have complex eigenvalues, direct differentiation through this quantity can lead to numerical instability during optimization. 
To avoid this difficulty, we instead introduce the symmetric matrix
\begin{equation}
{S} = \frac{{A} + {A}^{T}}{2},
\end{equation}
whose eigenvalues are guaranteed to be real and satisfy
\begin{equation}
\max\bigl( \mathrm{Re}(\lambda({A})) \bigr)
\le
\max\bigl( \lambda({S}) \bigr).
\end{equation}
Accordingly, a sufficient condition for Hurwitz stability is that the largest eigenvalue of ${S}$ be negative. 
In this work, we therefore define the stability loss as
\begin{equation}
\mathcal L_{\mathrm{stability}}(A) =\lambda_{\mathrm{stability}}
\mathrm{ReLU}\!\Big(
\max \big( \lambda({S}) \big)
\Big).
\end{equation}
The effect of the Hurwitz stability constraint on long-time population predictions is demonstrated in Section~\ref{sec:result}. However, while the Hurwitz constraint ensures that the system converges to some steady state, it does not guarantee that the converged state coincides with the physically correct steady state obtained from reference NLTE calculations. This limitation is addressed in the following subsection through an explicit steady-state constraint.

\subsubsection{Steady-State Consistency}

Under fixed temperature and density, the latent dynamics admit an equilibrium point
\begin{equation}
\mathbf{z}^\ast = -{A}^{-1}(\mathbf{a}+{B}\mathbf{u}^\ast),
\end{equation}
obtained by setting $d\mathbf{z}/dt=0$. If ${A}$ satisfies the Hurwitz stability condition, trajectories starting from arbitrary initial conditions converge to this equilibrium. Because the LaSDI model is formulated as an explicit latent ODE, this steady state can be computed analytically and mapped through the decoder to obtain the corresponding physical population $\mathbf{n}^\ast$. However, stability alone is not sufficient: while the Hurwitz condition guarantees convergence, it does not ensure that the converged equilibrium is the physically correct NLTE steady state. In practice, a model trained only with stability enforcement may still converge to a shifted or incorrect equilibrium. To address this issue, a steady-state loss is introduced to explicitly constrain the latent equilibrium to match the reference NLTE steady-state population under fixed plasma conditions.

In principle, such behavior could be learned implicitly from very long time-dependent simulations that include fully relaxed states. For TD-NLTE problems, however, this would substantially increase the number of time steps, the dataset size, and the training cost. Instead, steady-state information is incorporated directly as a separate physical constraint. The 185 time-dependent training trajectories contain about 74,000 population distributions in total, whereas only about 200 steady-state samples are used to impose the steady-state constraint. This provides an efficient way to enforce physically correct steady-state behavior without directly learning long-time simulation data.

Specifically, the steady state predicted by the DMDc latent dynamics is compared with the reference NLTE steady state, and the following loss is defined:
\begin{equation}
\mathcal{L}_{\mathrm{steady}} = \lambda_{\mathrm{steady}}(\lambda_{n}\| \mathbf{n}_{\mathrm{steady}} - \mathbf{n}^{\ast} \|^{2} + \lambda_{p}\| \mathbf{p}_{\mathrm{steady}} - \mathbf{p}^{\ast} \|^{2} + \lambda_{\bar{q}}\| \bar{q}_{\mathrm{steady}} - \bar{q}^{\ast} \|^{2}).
\end{equation}
Here, the superscript ${}^{\ast}$ denotes quantities reconstructed through the decoder from the latent steady state computed from the DMDc formulation. A weak constraint is employed that does not include a loss term on the w-scaled population, since empirically such a term was found to degrade steady-state reproduction performance. All steady-state samples are drawn from fixed temperature--density conditions within the range covered by the time-dependent training data. The effect of this constraint on the predicted equilibrium is presented in Section~\ref{sec:result}. Through this formulation, the LaSDI model is extended into a TD-NLTE surrogate that satisfies not only stability but also physically correct steady-state convergence.

\subsection{Physics-Based Regularization}

The w-scaling maps the full dynamic range of populations into [0,1], making low-population states visible to the optimizer. However, in the present autoencoder, the use of a sigmoid activation in the final decoder layer introduces an additional trade-off: when w-scaling is applied, large original-scale populations are mapped close to zero in w-space, which can reduce reconstruction accuracy in the dominant-population regime. To mitigate this effect, we further augment the training objective with an additional reconstruction penalty in the original population space so that both low-population and dominant-population regimes are simultaneously constrained. The resulting combined reconstruction loss is defined as
\begin{equation}
\begin{aligned}
\mathcal{L}_{\mathrm{rec}}=\| \mathbf{w} - \hat{\mathbf{w}} \|^{2}+\lambda_{n}\| \mathbf{n} - \hat{\mathbf{n} \|^{2}}.
\end{aligned}
\end{equation}
where $\hat{\cdot}$ denotes the autoencoder reconstruction, i.e., $\hat{\mathbf{w}} = \mathcal{G}_{\mathrm{dec}}(\mathcal{G}_{\mathrm{enc}}(\mathbf{w}))
$. By using this combined loss, the relative reconstruction error is reduced more uniformly across the full population range. In this study, we set $\lambda_n = 1$.

In addition to this representation-level regularization, we also include several auxiliary physical penalties based on macroscopic observables. Although the atomic population vector contains the most detailed microscopic information, total population conservation, the charge-state distribution $p(q)$, and the mean charge state $\bar{q}$ are more directly relevant to multiphysics coupling and physical interpretation. Small level-wise reconstruction errors can therefore accumulate into noticeable errors in these aggregated quantities, even when the microscopic reconstruction itself appears accurate. Accordingly, the final autoencoder loss is augmented with penalty terms enforcing total population conservation, charge-state consistency, and mean-charge-state consistency:
\begin{equation}
\begin{aligned}
\mathcal{L}_{\mathrm{AE}} = \mathcal{L}_{rec}+ \lambda_{\Sigma}\|\Sigma \mathbf{n}_{\mathrm{truth}} - \Sigma \mathbf{n}_{\mathrm{AE}} \|^{2}+ \lambda_{p}\| \mathbf{p}_{\mathrm{truth}} - \mathbf{p}_{\mathrm{AE}} \|^{2}+ \lambda_{\bar{q}}\| \bar{q}_{\mathrm{truth}} - \bar{q}_{\mathrm{AE}} \|^{2}.
\end{aligned}
\end{equation}
where we set $\lambda_{\Sigma}=10^{-3}$, $\lambda_{p}=10^{-2}$ and $\lambda_{\bar{q}}=10^{-4}$ empirically.
Here, the conservation term is particularly useful because the combination of w-scaling and sigmoid decoding does not automatically guarantee exact normalization of the reconstructed population fractions.

In practice, the standalone impact of these auxiliary regularization terms on the validation error is modest, typically at the level of only about 1$-$2\% in our experiments. Nevertheless, they provide a low-cost mechanism for improving physical consistency and reducing the risk of macroscopic inconsistency during training. Since $p(q)$ and $\bar{q}$ are obtained through simple summation and weighted averaging over the reconstructed populations, the additional computational overhead is also modest, increasing the training time by only about 10\% in our tests. For this reason, we retain these terms as lightweight physics-based regularization throughout the training pipeline. Combining all terms, the full training objective is
\begin{equation}
\begin{aligned}
\mathcal{L}_{\mathrm{total}} = \mathcal{L}_{\mathrm{AE}} + \mathcal{L}_{\mathrm{DMDc}} + \mathcal{L}_{\mathrm{stability}} + \mathcal{L}_{\mathrm{steady}},
\end{aligned}
\end{equation}
where each term is summarized in Table~\ref{tab:loss_summary}. All weights are gradually increased during training to avoid destabilizing the autoencoder in early epochs.

\begin{table}[htbp]
\centering
\caption{Summary of the loss terms in the total training objective $\mathcal{L}_{\mathrm{total}}$.}
\label{tab:loss_summary}
\begin{tabular}{l l}
\hline
Category & Expression \\
\hline
$\mathcal{L}_{\mathrm{AE}}$
  & $\|\mathbf{w} - \hat{\mathbf{w}}\|^{2}
   + \lambda_{n}\|\mathbf{n} - \hat{\mathbf{n}}\|^{2}
   + \lambda_{\Sigma}\|\sum_i n_{i} - \sum_i \hat{n}_{i}\|^{2}
   + \lambda_{p}\|\mathbf{p} - \hat{\mathbf{p}}\|^{2}
   + \lambda_{\bar{q}}\|\bar{q} - \hat{\bar{q}}\|^{2}$ \\[6pt]
$\mathcal{L}_{\mathrm{DMDc}}$
  & $\lambda_{\mathrm{DMDc}}\|\dot{\mathbf{z}} - (\mathbf{a} + A\mathbf{z} + B\mathbf{u})\|^{2}$ \\[6pt]
$\mathcal{L}_{\mathrm{stability}}$
  & $\lambda_{\mathrm{stab}}\,\mathrm{ReLU}\!\big(\lambda_{\max}(S)\big)$ \\[6pt]
$\mathcal{L}_{\mathrm{steady}}$
  & $\lambda_{steady}(\lambda_{n}\|\mathbf{n}_{\mathrm{steady}} - \mathbf{n}^{\ast}\|^{2}
   + \lambda_{p}\|\mathbf{p}_{\mathrm{steady}} - \mathbf{p}^{\ast}\|^{2}
   + \lambda_{\bar{q}}\|\bar{q}_{\mathrm{steady}} - \bar{q}^{\ast}\|^{2})$ \\[6pt]
\hline
\end{tabular}
\end{table}

\section{Problem Setup and Data Generation}
\label{sec:problem_setup_data}

\subsection{Test Problem and Two-Stage Simulation Workflow}

To construct training and validation data for the proposed NLTE LaSDI model, we consider a proof-of-concept test problem motivated by laser-heated tin plasmas relevant to EUV lithography \cite{shi2023enhanced, sohn2025characterization}. The reference data are generated through a two-stage simulation workflow. In the first stage, radiation--hydrodynamic simulations are used to produce time-dependent temperature--density trajectories representative of the evolving plasma conditions. In the second stage, these trajectories are used as inputs to time-dependent NLTE calculations, which generate the corresponding atomic population histories.

The radiation--hydrodynamic simulations are performed using the Eulerian, multidimensional code FLASH \cite{fryxell2000flash}. To extract physically consistent local thermodynamic histories from the evolving plasma, we use the Lagrangian particle capability available in FLASH. This allows us to track moving plasma elements and record their temperature and density as functions of time. The resulting $T$--$\rho$ trajectories serve as the driving input for the subsequent atomic kinetics calculations.

Time-dependent NLTE calculations are then carried out using SCFLY, a widely used collisional--radiative code based on the super-configuration approach \cite{chung2017atomic, cho2024nonequilibrium}. In this work, SCFLY is used solely as a data generator for atomic population dynamics. By focusing on population evolution rather than radiative outputs, we isolate the kinetic aspects of the NLTE problem and assess how accurately the proposed surrogate model can reproduce the underlying dynamics. For the present tin test problem, the SCFLY atomic model contains 1,583 super-configurations, and the output of the second stage is a collection of time-resolved high-dimensional population vectors that serve as the reference data for training and evaluating the reduced-order surrogate model.

\subsection{Dataset Construction and Simplifications for the Proof-of-Concept Study}

\begin{figure}[htbp]
    \centering
    \includegraphics[width=\linewidth]{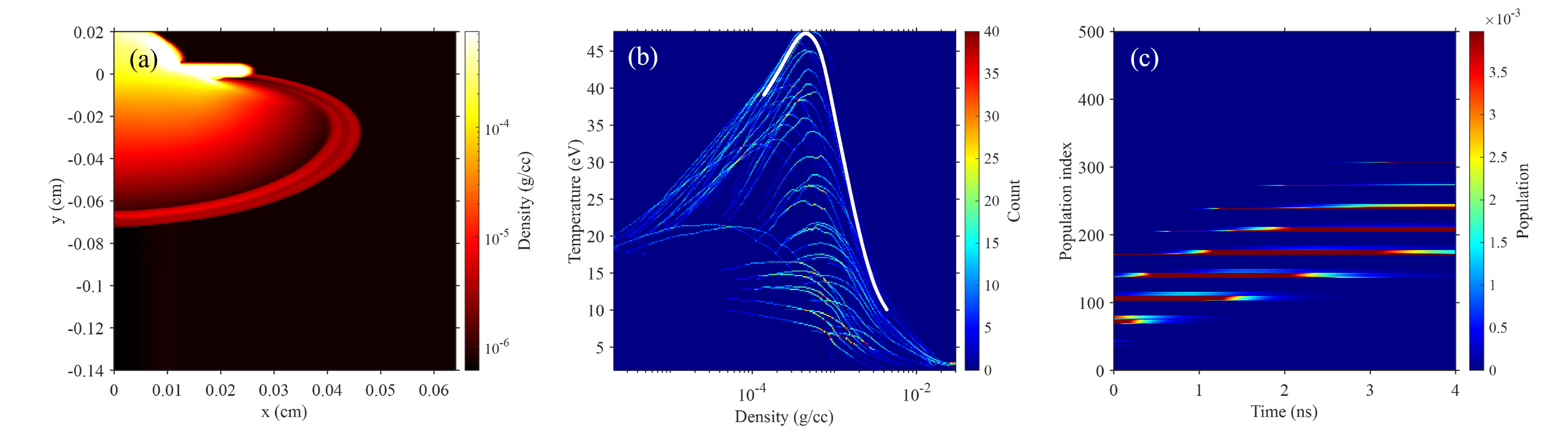}
    \caption{Overview of the data generation workflow. (a) Radiation--hydrodynamic simulation of a laser-heated tin plasma using FLASH, showing the spatial distribution of density. (b) Temperature--density trajectories of Lagrangian particles extracted from the FLASH simulation, with the color map indicating the sample count across the parameter space. (c) TD--NLTE population distributions generated by SCFLY using the temperature--density trajectories from (b) as input, shown for the representative trajectory highlighted in (b).}
    \label{fig:Trho_distribution}
\end{figure}

Using the FLASH--SCFLY workflow, we generate 185 distinct temperature--density trajectories, each spanning 4~ns with a temporal resolution of $\Delta t = 0.01$~ns; 147 trajectories (approximately 80\%) are used for training and the remaining 38 for validation. These temperature--density histories are then supplied to SCFLY to generate time-resolved NLTE population trajectories, represented for the present tin test problem by a 1,583-dimensional super-configuration state vector. In addition, 200 steady-state NLTE samples are generated for the steady-state constraint by uniformly sampling the temperature--density domain covered by the trajectory dataset, with temperatures ranging from 1 to 50~eV and densities from $10^{-5}$ to $10^{-2}$~g/cc.Figure~\ref{fig:Trho_distribution} illustrates the three-stage data generation workflow: radiation--hydrodynamic simulations with FLASH provide temperature--density trajectories, which are then supplied to SCFLY to generate TD-NLTE population distributions.

To isolate the core methodological question of this work, several simplifying assumptions are adopted. First, all NLTE calculations are performed under optically thin conditions, so radiative-transport feedback is neglected and population evolution is driven solely by the prescribed temperature and density histories. This allows the analysis to focus on the stiff atomic kinetics and long-time behavior most relevant to stability and steady-state consistency. Second, the NLTE reference calculations are based on the SCFLY super-configuration model rather than a more detailed atomic representation. Although this reduced model may not capture all fine spectral features, it provides a suitable and computationally tractable benchmark for constructing and validating a reduced-order surrogate for time-dependent atomic population dynamics.

Taken together, these choices define a controlled proof-of-concept setting in which the proposed surrogate can be tested against physically meaningful NLTE kinetics while avoiding additional complications associated with radiation transport and highly detailed atomic structure. Extensions to more general NLTE problems with explicit radiation coupling and higher-fidelity atomic models are left for future work.

\section{Results}
The SCFLY super-configuration model defines the NLTE atomic population as a 1,583-dimensional vector, which is compressed by the proposed model into only three latent variables, corresponding to a dimensionality reduction factor of approximately 521. The autoencoder uses five hidden layers in both the encoder and decoder, with layer sizes of [800, 400, 200, 100, 20]. Mish activation functions are used in the hidden layers, and a sigmoid activation is applied in the final decoder layer to ensure that the w-scaled population values remain within $[0,1]$. The model is trained end-to-end for 30,000 epochs on 147 trajectories, with the remaining 38 trajectories reserved for validation. To improve robustness, Gaussian noise ($\sigma = 0.1$) is injected into the latent variables during training, and the weights of the macroscopic, steady-state, and stability constraints are gradually increased using a ramp-up strategy.
\label{sec:result}

\subsection{Identified Latent Dynamics Coefficients}

The pLaSDI training produces a three-dimensional latent dynamical system of the form~\eqref{eq:DMDc} with the following globally identified coefficients:
\begin{equation}
\mathbf{a} = \begin{pmatrix} 43.42 \\ 53.72 \\ -126.11 \end{pmatrix}, \quad
A = \begin{pmatrix}
-1.42 & -1.61 & -4.32 \\
 2.41 & -3.95 &  8.92 \\
 2.04 &  8.55 & -82.97
\end{pmatrix}, \quad
B = \begin{pmatrix}
 25.46 & -28.35 \\
-63.64 &  58.62 \\
798.53 & -269.38
\end{pmatrix}.
\end{equation}
The real part of eigenvalues of $A$ are $\lambda_1 =\lambda_2\approx -2.25$ and $\lambda_3 \approx -83.84$. This confirmes that the Hurwitz stability condition is satisfied. The wide spread in eigenvalue magnitudes reflects the multi-timescale relaxation structure inherent in NLTE kinetics. The large magnitude of the $B$ coefficients indicates that this latent variable is sensitive to changes in plasma conditions.

\subsection{Reconstruction of Population Dynamics and Macroscopic Observables}

\begin{figure}[htbp]
    \centering
    \includegraphics[width=\linewidth]{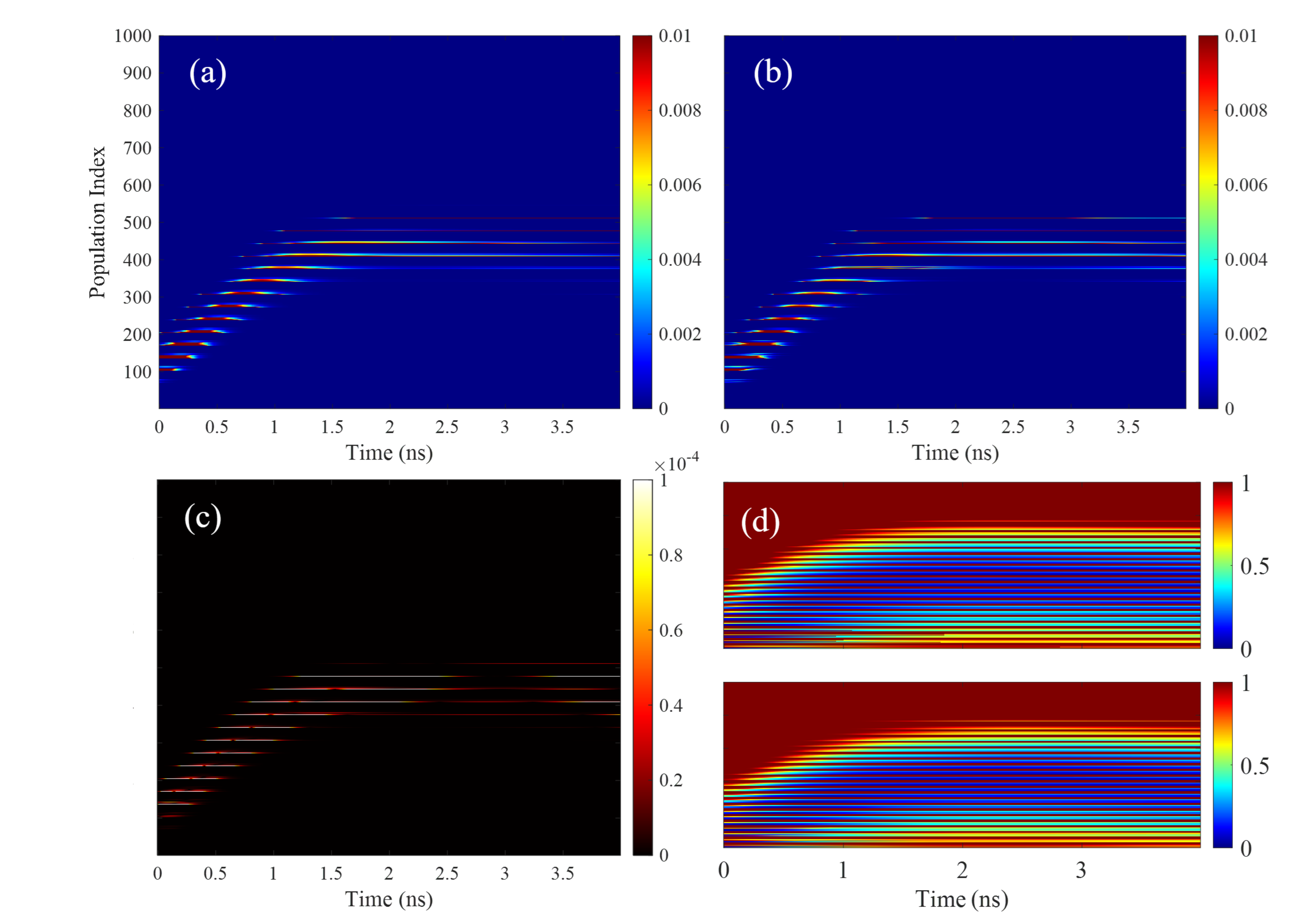}
    \caption{Comparison of microscopic population predictions for a representative validation case. (a) Fractional population from the SCFLY reference, (b) fractional population from the pLaSDI prediction, and (c) their pointwise absolute error. (d) shows the corresponding w-scaled populations from SCFLY and pLaSDI, highlighting agreement in the low-population regime. All panels share the same y-axis range.}
    \label{fig:pop_comparison}
\end{figure}

We first assess whether the pLaSDI model can accurately reproduce the microscopic atomic population evolution and the macroscopic quantities derived from it. Figure~\ref{fig:pop_comparison} compares the SCFLY reference and the pLaSDI prediction for a representative validation case. The primary comparison is carried out in the original fractional population space, since it directly shows how well the physically important dominant population structures are preserved. In addition, the comparison in the w-scaled population space is included to assess reconstruction performance in the low-population regime, which is not easily visible in the original population space. The w-scaled results provide complementary evidence of the model’s ability to represent small-population states. 

In the original fractional population space, pLaSDI reproduces the dominant time-dependent population structures well, with errors confined to limited regions. This indicates that the surrogate preserves the physically important main population dynamics. At the same time, the close visual agreement in the w-scaled space shows that reconstruction quality is maintained even in the low-population regime. Averaged over all validation trajectories, the root-mean-square error (RMSE) is approximately $2.9\times 10^{-3}$ for the fractional populations and $2.8\times 10^{-2}$ for the w-scaled populations. These results show that the proposed model preserves the essential temporal structure of the NLTE solution while effectively representing population distributions over a very wide dynamic range, including both dominant and small-population states.

\begin{figure}[htbp]
    \centering
    \includegraphics[width=\linewidth]{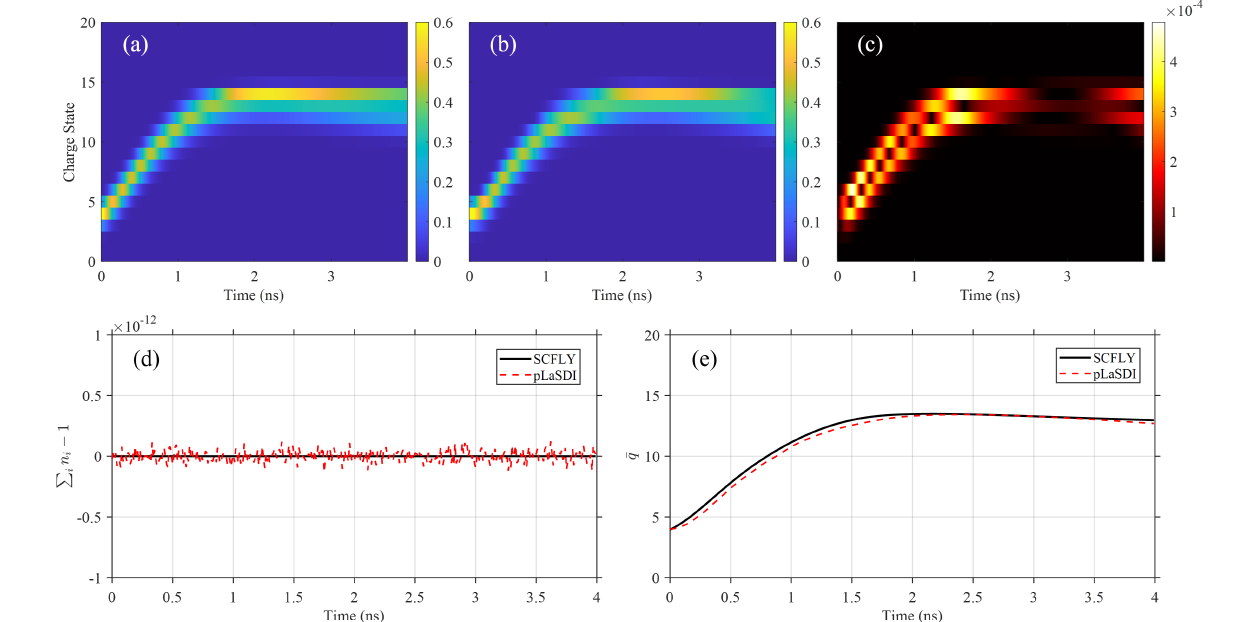}
    \caption{Comparison of macroscopic observables for a representative validation case. (a--c) Charge-state distribution: (a) SCFLY reference, (b) pLaSDI prediction, (c) pointwise absolute error. (d) Population conservation ($\sum_i n_i - 1$); the deviation remains at the $10^{-13}$ level. (e) Mean charge state $\bar{q}$.}
    \label{fig:macro_comparison}
\end{figure}

To assess whether this microscopic agreement translates into physically meaningful macroscopic behavior, Figure~\ref{fig:macro_comparison} compares the predicted CSD, population conservation, and mean charge state $\bar{q}$ with the SCFLY reference.
The predicted CSD agrees closely with the SCFLY reference across the full time region, with an RMSE of approximately $1.6\times 10^{-2}$ averaged over all validation trajectories. The reconstructed populations satisfy conservation to within $\mathcal{O}(10^{-13})$, and the mean charge state is tracked with high accuracy, with a trajectory-averaged relative error in $\bar{q}$ of approximately 1.92\% for the training set and 1.98\% for the validation set. Since the CSD and $\bar{q}$ are closely tied to radiative and thermodynamic properties, these results show that the surrogate preserves physically meaningful structure beyond pointwise population fitting.

\subsection{Role of Physics-informed Constraints and Steady-state Generalization}

\begin{figure}[htbp]
\centering
\includegraphics[width=0.8\linewidth]{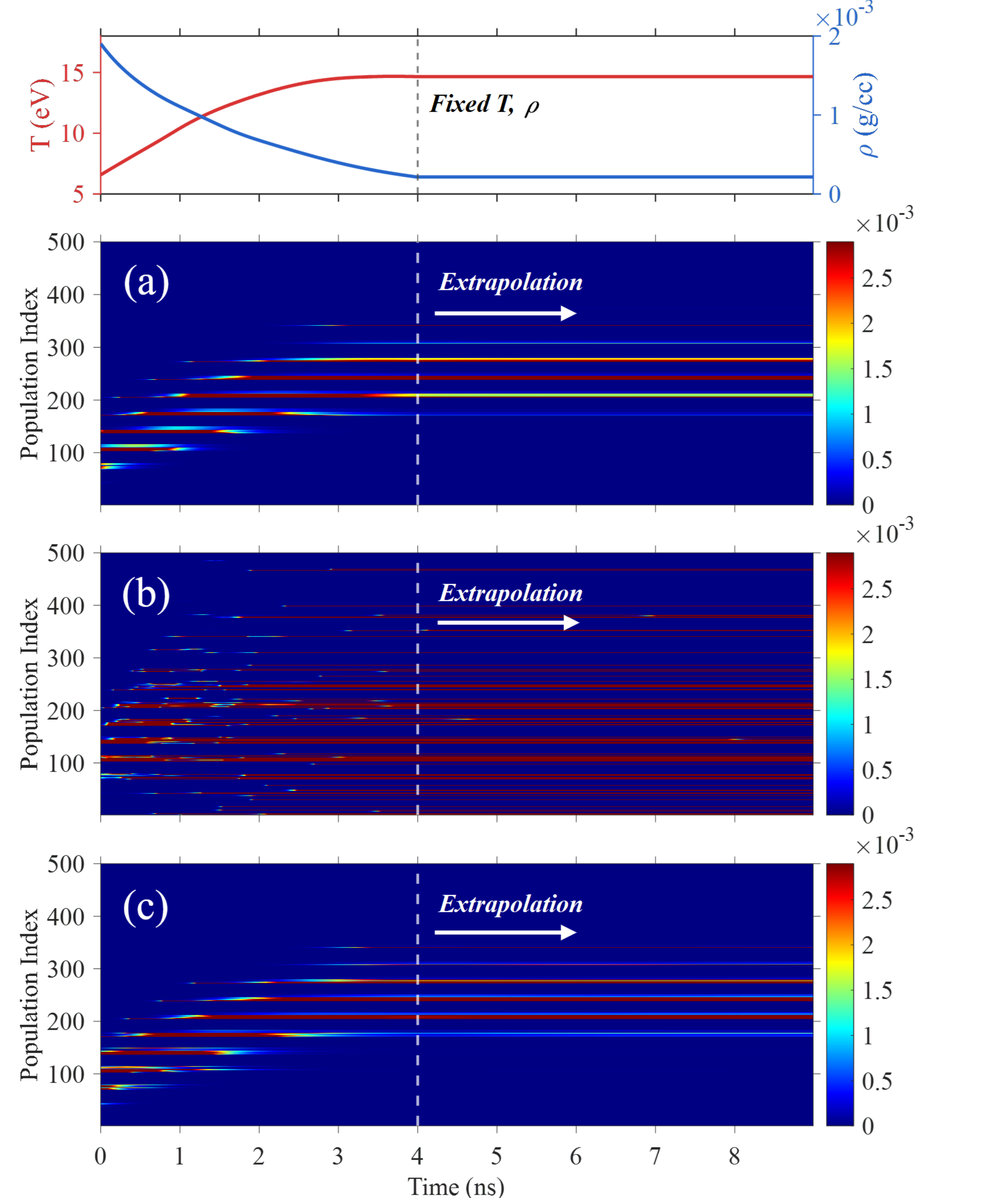}
\caption{Effect of the Hurwitz stability constraint on long-time population predictions. (a) Reference SCFLY population distribution up to 4 ns; the inset shows the prescribed temperature and density histories, which are held fixed beyond 4 ns. (b) Without the Hurwitz constraint; the learned operator $A$ has large eigenvalues with positive real parts (severely violating the Hurwitz condition), and the predicted populations diverge significantly
even within the training interval. (c) With the Hurwitz constraint, the populations remain bounded and converge to a stable distribution consistent with the reference.}
\label{fig:hurwitz_comparison}
\end{figure}

This subsection examines the role of the Hurwitz stability constraint in the learned latent dynamics. Figure~\ref{fig:hurwitz_comparison}(a) shows a representative SCFLY population trajectory from the dataset over the 4~ns training interval. To assess long-time stability, the temperature and density are held fixed beyond 4~ns, and the learned latent dynamics are integrated forward in time. Without the Hurwitz constraint [Figure~\ref{fig:hurwitz_comparison}(b)], the predicted populations diverge not only during the extrapolation period but even within the training window when the latent ODE is integrated forward. This behavior does not imply poor agreement with the training snapshots themselves; rather, it reflects instability in the learned dynamical system during time integration. The DMDc fitting residuals are comparable between cases (b) and (c), indicating that this divergence is not caused by poor local derivative fitting, but by the unstable eigenvalue structure of the learned operator.

Specifically, the DMDc coefficients are obtained by pointwise regression on time derivatives, which inevitably introduces small fitting errors at each snapshot. When the learned operator $A$ possesses eigenvalues with large positive real parts, as is the case in (b) where the Hurwitz condition is severely violated, any small deviation from the nominal latent trajectory is amplified at every integration step, causing rapid divergence even within the training window. By contrast, when the Hurwitz constraint is enforced [Figure~\ref{fig:hurwitz_comparison}(c)], the latent dynamics act as a stable attractor, continuously driving integration errors back toward the nominal trajectory so that the predicted populations remain bounded and converge stably, reproducing the reference population structure even beyond the training window. It should be noted, however, that the Hurwitz condition restores the trajectory toward the equilibrium of the \emph{learned} dynamical system; residual long-time errors therefore reflect the approximation gap in the identified DMDc coefficients rather than dynamical instability.

\begin{figure}[htbp]
    \centering
    \includegraphics[width=\linewidth]{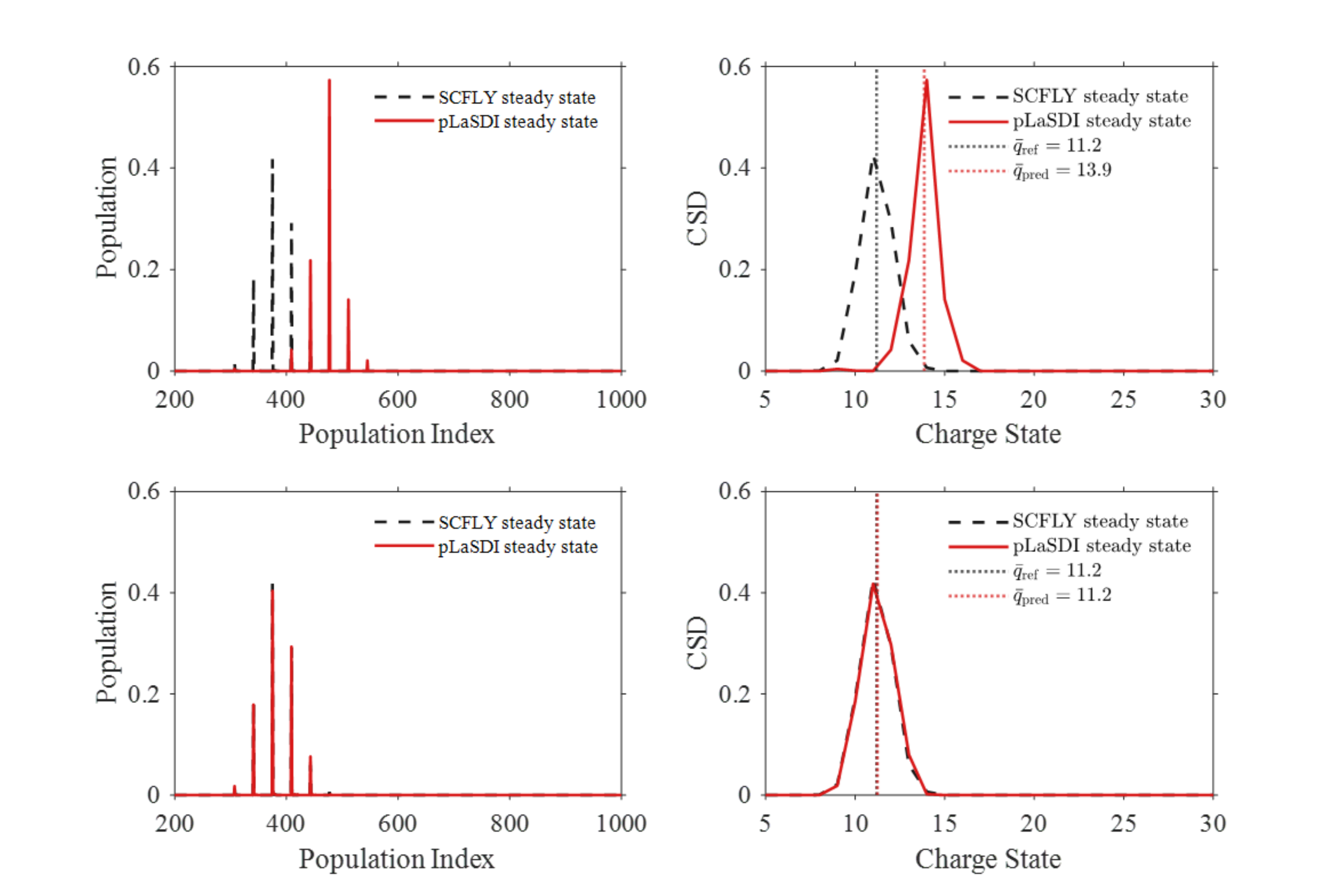}
    \caption{Effect of the steady state constraint on the predicted equilibrium population and charge state distribution at $T=45$ eV and $\rho=10^{-5}$ g/cc. (a, b) Without the constraint, the predicted population and CSD are visibly shifted from the reference. (c, d) With the constraint, the predicted distributions closely overlap the reference. The population panels are zoomed in to super-configuration indices 200--1000 (of 1--1,583), and the CSD panels to charge states 5--30 (of 0--50), to highlight the dominant features.}
    \label{fig:ss_comparison}
\end{figure}

The steady-state constraint plays a complementary role to the Hurwitz stability condition. While the Hurwitz condition guarantees bounded convergence, it does not ensure that the asymptotic state coincides with the physically correct NLTE equilibrium. As shown in Figure~\ref{fig:ss_comparison}(a,b), the latent population may converge to a stable equilibrium that remains shifted from the reference NLTE steady state. When the steady-state constraint is included during training, however, the predicted population distribution closely matches the reference steady state, and the mean charge state is also reproduced with high accuracy [Figure~\ref{fig:ss_comparison}(c,d)]. Together, these two constraints enable the LaSDI model to move beyond short-time trajectory fitting and serve as a physically reliable surrogate for long-time NLTE simulations.

\begin{figure}[htbp]
    \centering
    \includegraphics[width=\linewidth]{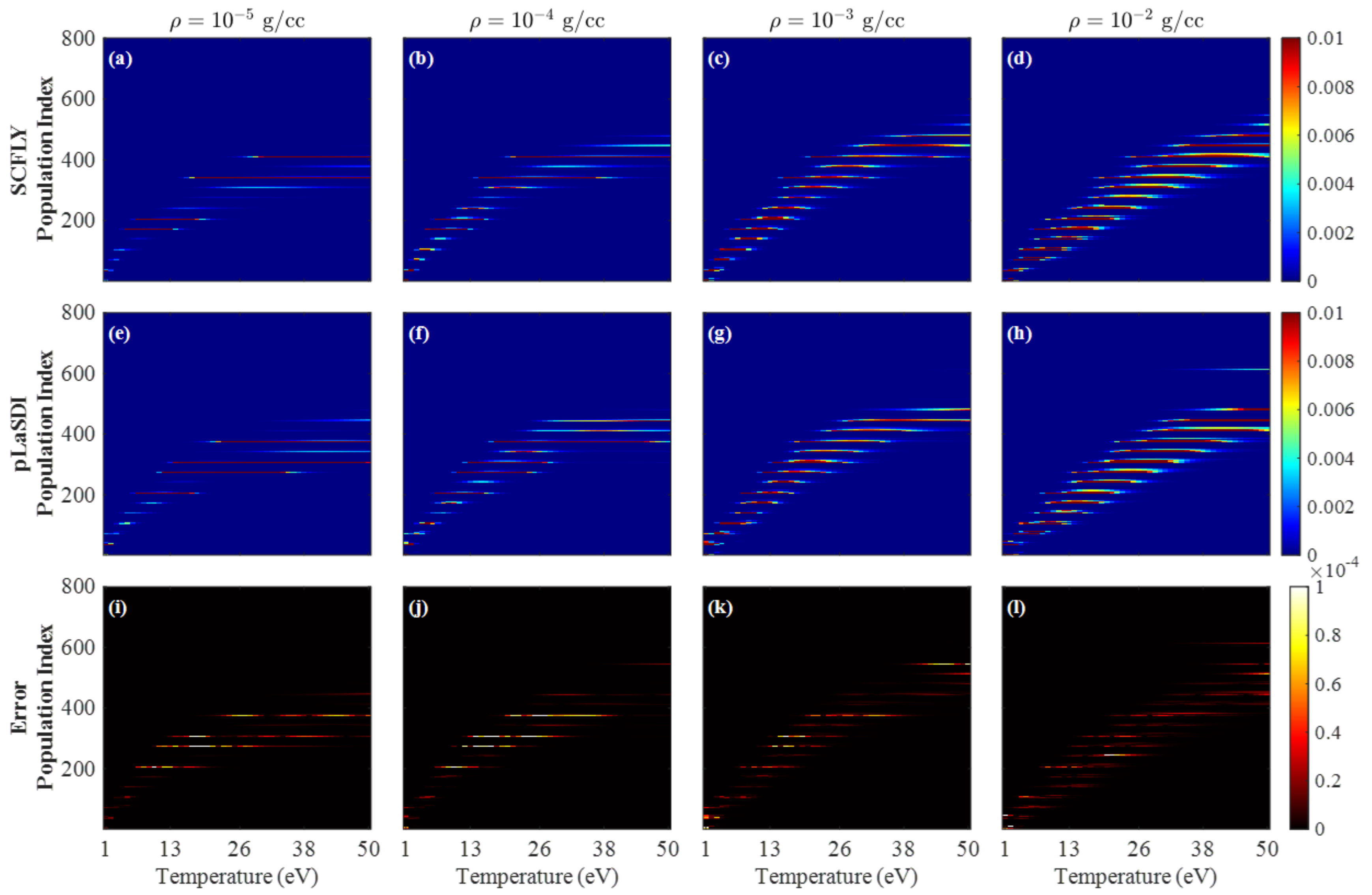}
    \caption{Steady-state population predictions from the analytically computed latent equilibrium $\mathbf{z}^\ast = -A^{-1}(\mathbf{a} + B\mathbf{u}^\ast)$, compared with SCFLY reference solutions across four densities ($10^{-5}$, $10^{-4}$, $10^{-3}$, $10^{-2}$~g/cc). Within each density block, temperature increases from 1 to 50~eV. (a--d) SCFLY reference, (e--h) pLaSDI prediction, (i--l) pointwise absolute error. Columns correspond to increasing density from left to right.}
    \label{fig:ss_heatmap}
\end{figure}

We further assess the quality of the learned equilibrium structure across a broad range of plasma conditions. Figures~\ref{fig:ss_heatmap} compares the steady-state population from the analytically computed latent equilibrium,
\[
\mathbf{z}^{\ast} = -A^{-1}(\mathbf{a} + B\mathbf{u}^{\ast}),
\]
with the reference SCFLY steady-state solutions over temperatures from 1 to 50~eV at four different densities ($10^{-5}$, $10^{-4}$, $10^{-3}$, $10^{-2}$~g/cc).

Across all four densities, the predicted steady-state population and charge-state distributions remain in close agreement with the reference solutions throughout the full temperature range, and the mean charge state $\bar{q}$ is reproduced with high fidelity at all densities. Quantitatively, the RMSE across all steady-state samples is $5.7\times10^{-4}$ for fractional populations, $3.2\times10^{-3}$ for the CSD, and $4.5\times10^{-2}$ for $\bar{Z}$. This is notable because only approximately 200 steady-state samples are used to impose the steady-state constraint. The results therefore suggest that the latent model has learned a physically meaningful equilibrium structure that generalizes over the temperature--density domain covered by the dataset.

\subsection{Computational Efficiency}

Finally, we assess the computational advantage of the surrogate model. Conventional SCFLY-based time-dependent NLTE calculations typically require computation times ranging from several tens of minutes to several hours, depending on the plasma conditions and trajectory. In contrast, the trained LaSDI model evaluates a full time-dependent population trajectory in less than approximately 0.038~s.

This corresponds to a speedup of approximately $5 \times 10^{4}$--$10^{5}$, depending on the problem setup. Such acceleration is substantial for applications in which NLTE calculations are repeatedly invoked, including radiation--hydrodynamic simulations, parameter scans, and optimization workflows. Combined with the demonstrated stability and steady-state fidelity, this performance indicates that the proposed physics-informed LaSDI model is not only accurate, but also practically viable as a fast surrogate for time-dependent NLTE atomic kinetics.

\section{Conclusion}
\label{sec:conclusion}

In this study, we have presented a physics-informed LaSDI-based surrogate model for time-dependent NLTE atomic kinetics calculations.The proposed model compresses high-dimensional atomic population dynamics into a small number of latent variables using a nonlinear autoencoder, and describes the temporal evolution in the latent space through explicit differential equations. A key design principle is the adoption of an explicit latent ODE--whose analytical structure allows the stability and steady-state properties of the learned dynamics to be characterized and constrained \emph{a priori}.

The ablation studies presented in this work highlight a broader lesson for surrogate modeling of stiff dynamical systems: data-driven fitting of time derivatives (e.g., via DMDc) can achieve low pointwise residuals while producing an operator whose eigenvalue structure leads to catastrophic failure upon time integration. The Hurwitz stability and steady-state consistency constraints introduced here address this gap by enforcing global properties of the dynamical system that cannot be inferred from local derivative matching alone. This physics-informed approach to latent dynamics design---constraining not only what the model fits but how it behaves---is, in our view, essential for constructing reliable surrogates for any stiff kinetic system, beyond NLTE applications.

As a result, the proposed model compresses the original 1,583-dimensional atomic population space into a three-dimensional latent space (a $521\times$ reduction), while stably reproducing key physical quantities such as time-dependent population evolution, charge state distributions, and mean charge states. The mean charge state is reproduced with relative errors of approximately 2\%, and the model achieves speedups of $5 \times 10^{4}$--$10^{5}$ relative to SCFLY calculations. Moreover, under long-term extrapolation conditions, the predicted population distributions converge to physically correct steady states.

Several limitations of the current study should be noted. This work considers optically thin conditions and employs a super-configuration-based NLTE model, without accounting for radiation transport effects or more detailed atomic structure models. The current latent dimension ($N_z = 3$) achieves high fidelity in macroscopic observables but leaves room for improvement in individual level-population accuracy; exploring larger latent dimensions to improve level-wise reconstruction is a natural next step. In addition, this work focuses on validating the accuracy and stability of atomic population dynamics, and therefore does not include explicit spectral calculations. Nevertheless, the high level of agreement observed in population distributions and charge state distributions suggests that the proposed framework is likely to yield favorable performance in future spectral and opacity calculations.

Future work will extend the present framework in several directions. These include incorporating radiation transport effects and applying the model to more detailed atomic structure data beyond the super-configuration representation. Another promising direction is to replace the current least-squares-based DMDc identification with alternative approaches---such as neural-network-based latent rate operators or architecturally constrained dynamics models---that can embed stability properties into the mathematical structure of the learned system itself, rather than enforcing them through auxiliary penalty terms. Finally, integrating this surrogate into radiation--hydrodynamic simulation workflows will be explored to assess its practical impact on end-to-end computational efficiency.

\section{Acknowledgments}
This work was performed under the auspices of the U.S. Department of Energy by Lawrence Livermore National Laboratory under Contract No. DE-AC52-07NA27344. This research was supported by the Institute for Basic Science (IBS- R038-D1) and the National Research Foundation of Korea (NRF) grant funded by the Korean government (RS-2023-00218180, RS-2025-00516264).

\bibliographystyle{unsrt}
\bibliography{mybib} 

\end{document}